\documentstyle[11pt]{article}
\begin{document}

\begin{center}


{\Large\bf Superconformal Anomaly from AdS/CFT Correspondence}
\renewcommand{\thefootnote}{\dagger}\footnote{Contribution
to `` Symmetries in Gravity and Field Theory " conference for
Professor A. de Azcarraga's 60th birthday, June 2003, Salmanca,
Spain}

\vspace{5mm}


{\large M. Chaichian}
and {\large W.F. Chen}

\vspace{2mm}

High Energy Physics Division, Department of Physical Sciences,
University of Helsinki \\
 and Helsinki Institute of Physics, FIN-00014,  Helsinki,
Finland

\end{center}


\begin{abstract}
\noindent  For a classical superconformal gauge theory in a
conformal supergravity background, its chiral R-symmetry anomaly,
Weyl anomaly and super-Weyl anomaly constitute a supermultiplet.
We review how these anomalies arise from the five-dimensional
gauged supergravity in terms of the AdS/CFT correspondence at the
gravity level. The holographic production of this full
superconformal anomaly multiplet provides a support and test to
the celebrated AdS/CFT conjecture.
\end{abstract}

\section{Introduction}

The discovery of a $D$-brane as a fundamental dynamical object
carrying $R-R$ charge has played a crucial role in establishing a
web of dualities among five superstring theories and unifying them
into a single M-theory \cite{pol}. On one hand, combining other
 types of branes such as the Neveu-Schwarz (NS) solitonic
 five-brane and orientifold plane, a supersymmetric gauge
theory defined on the world volume of $Dp$-branes can be
constructed from an elaborated setting of a brane configuration in
a weakly type II string theory \cite{giku}. On the other hand, a
stack of $Dp$-branes can modify the space-time background of the
strong coupled type II string theory and arise as  the brane
solution to the low-energy effective theory of type II string,
i.e., type II ($A$ or $B$) supergravity. These two distinct
features of a D-brane at both strong and weakly superstring theory
have led to the celebrated  AdS/CFT correspondence conjecture
proposed by Maldacena \cite{mald}.

  The original AdS/CFT
correspondence conjecture \cite{mald}
  states that  the type IIB string theory compactified
  on $AdS_5\times S^5$
  theory with $N$ units of $R-R$ flux on $S^5$ describes the same physics
  as ${\cal N}=4$ $SU(N)$ supersymmetric Yang-Mills theory.
  The explicit definition was further clarified and generalized
   as the following \cite{gkp,witt1}. Given
 the type II superstring theory in the background
$AdS_{d+1}\times X^{9-d}$,
 with $X^{9-d}$ being a compact Einstein manifold,
there should exist a one-to-one correspondence between a string
state a supergravity field in the $AdS_{d+1}$ bulk and a gauge
invariant operator of the conformal field theory defined on the
boundary of $AdS_{d+1}$ space-time. Concretely, the partition
function $Z_{\rm String}\left[\phi_0\right]$ of the type II
superstring theory as a functional of the boundary value
$\phi_0(x)$ of a bulk field $\phi (x,r)$  should equal the
generating functional of the correlation function $Z_{\rm CFT}$ of
the gauge invariant operator of the conformal conformal field
theory with the external source $\phi_0(x)$  provided by the
boundary value of the bulk field $\phi (x,r)$,
 \begin{eqnarray}
  Z_{\rm String}\left[\phi_0\right]
   &=& Z_{\rm CFT}\left[\phi_{0}\right], \nonumber\\
\left.Z_{\rm String}\left[\phi_0\right]\right| &=& \int_{\phi
(x,0)=\phi_{0}(x)}{\cal D}\phi (x,r)
\exp\left(-S[\phi (x,r)]\right), \nonumber\\
Z_{\rm CFT}\left[\phi_{0}\right]&=& \left\langle \exp\int_{M^d}
d^dx
{\cal O} (x)  \phi_{0}(x)\right\rangle \nonumber\\
&=&\sum_n\frac{1}{n!}\int \prod_{i=1}^n d^d x_i \left\langle {\cal
O}_1 (x_1)\cdots {\cal O}_n (x_n) \right\rangle \phi_{0} (x_1)
\cdots \phi_{0} (x_n)\nonumber\\
&{\equiv}& \exp\left(-\Gamma_{\rm CFT} [\phi_{0}]\right).
\label{acc1}
\end{eqnarray}
In above equation, ${\cal O}$ represent certain composite operator
in the superconformal field theory such as the energy-momentum
tensor and chiral $R$-symmetry current etc., $\phi_0 $ are the
corresponding background fields such as the gravitational and
gauge  fields etc. coming from the boundary value of the
corresponding bulk field, and  $\Gamma_{\rm CFT} [\phi_{0}]$ is
 the quantum effective action describing
the composite operators  interacting with
 background field $\phi_{0}$.

At low-energy the string effect can be neglected. The  partition
function of the type IIB superstring can be evaluated as the
exponential of the type IIB supergravity action in a on-shell
field configuration $\phi^{\rm cl}[\phi^{0}]$ with the boundary
value $\phi_{0}(x)$, i.e.,
\begin{eqnarray}
Z_{\rm String}\left[\phi_0\right]
 =\exp\left(-S_{\rm SUGRA}[\phi^{\rm cl}[\phi_{0}]]\right) \,.
\label{acc3}
\end{eqnarray}
A comparison between Eq.\,(\ref{acc1}) and Eqs.\,(\ref{acc3})
immediately shows that the large-$N$ quantum effective action of
the $d$-dimensional conformal
 field theory in the background provided by $\phi_0$
 can be approximately
equal to the on-shell classical action of  $AdS_{d+1}$
supergravity with non-empty boundary,
\begin{eqnarray}
\Gamma_{\rm CFT}[\phi_{0}]=S_{\rm SUGRA}[\phi^{\rm
cl}[\phi_{0}]]=\int d^d x \phi_0 (x) \left\langle {\cal
O}\right\rangle. \label{acc4}
\end{eqnarray}

Let us see how the five-dimensional gauged supergravities
\cite{gst,awada,guna2} arises in the AdS/CFT correspondence
\cite{ferr}.  The $AdS_5\times S^5$ background comes from the near
horizon limit of three-brane solution of type IIB supergravity
\cite{agmo}. Therefore, in the $AdS_5\times S^5$ background, the
spontaneous compactification on $S^5$ of the type IIB
supergravity occurs \cite{freu}. With the assumption that there
exists a consistent nonlinear truncation of the massless modes
from the whole Kluza-Klein spectrum of the type IIB supergravity
compactified on $S^5$ \cite{marcus,kim}, the resultant theory
should be the $SO(6)(\cong SU(4))$ gauged ${\cal N}=8$ $AdS_5$
supergravity since the isometry group $SO(6)$ of $S^5$ becomes the
gauge group of the compactified theory and the $AdS_5\times S^5$
background preserves all the supersymmetries of type IIB
supergravity \cite{guna2}. Furthermore, if the background for the
type IIB supergravity is  $AdS_5 \times X^5$ with $X^5$ being an
Einstein manifold rather than $S^5$ such as $T^{1,1}=(SU(2)\times
SU(2))/U(1)$ or certain orbifold, then the number of preserved
supersymmetries in the compactified $AdS_5$ supergravity is
reduced \cite{roman2,kw2}. One can thus obtain the gauged ${\cal
N}=2,4$ $AdS_5$ supergravity in five dimensions, and their dual
field theories should
 be $N=1,2$ supersymmetric gauge theories \cite{ferr}.
A supersymmetric gauge theory with lower supersymmetries is  not a
conformal invariant theory since its beta function usually does
not vanish. However, it was shown that the beta function a
supersymmetric gauge theory has the zero point, at which the
(super)conformal invariance can arise \cite{shif,seib}. The
$AdS/CFT$ correspondence between the ${\cal N}=2,4$ gauged
supergravities in five dimensions and  ${\cal N}=1,2$
supersymmertric gauge theories can thus be established
\cite{ferr}.

Eq.\,(\ref{acc4}) shows clearly that according the AdS/CFT
correspondence
 a quantum effective action
describing a superconformal gauge theory in an external
supergravity background can be identified with the on-shell action
of the gauged supergravity with non-trivial boundary data. Thus
the various anomalies at the leading-order of large-$N$ expansion
of a four-dimensional supersymetric gauge theory should be
extracted from the $AdS_5$ gauged supergravity. Specifically, the
AdS/CFT correspondence exchange strong and weak couplings and vice
versa, and the anomalies are independent to the couplings, the
production of the  superconformal anomaly from the gauged
supergravity will provide a supportand and test to the $AdS/CFT$
correspondence at the supergravity level.

This short review is outlined as the following. In Sect.\,II,  we
shall briefly introduce a classical superconformal gauge theory in
a conformal supergravity background and explain how a
superconformal anomaly multiplet arises. In Sect.\,III we shall
explain the structure of the gauged supergravity in five
dimensions and emphasize why the superconformal anomaly multiplet
can be reproduced from the gauged supergravity in terms of the
AdS/CFT correspondence. Sect.\,IV contains a systematics review on
how chiral-, Weyl- and super-Weyl anomalies arise from the
classical five-dimensional gauged supergravity. Sect.\,V is a
brief summary.

\section{Superconformal Anomaly Multiplet in External
Conformal Supergravity Background}

We focus on  a general ${\cal N}=1$ four-dimensional
supersymmetric $SU(N)$ gauge theory,
 its conserved currents including the energy-momentum
tensor $\theta^{\mu\nu}$, the supersymmetry current $s^{\mu}$ and
the chiral (or equivalently axial vector) R-current $j^{\mu}$
constitute a supermultiplet due to the supersymmetry
\cite{ilio,anm},
\begin{eqnarray}
\partial_\mu T^{\mu\nu}=\partial_\mu s^\mu=\partial_\mu
j^{\mu}=0.
\end{eqnarray}
If these currents satisfy further algebraic constraints,
$T^{\mu}_{~\mu}=\gamma_\mu s^\mu=0$, the Poincar\'{e}
supersymmetry will be promoted to a superconformal symmetry since
one can construct three more conserved currents,
\begin{eqnarray}
d^\mu {\equiv} x_\nu T^{\nu\mu}, ~ k_{\mu\nu}{\equiv} 2 x_\nu
x^\rho T_{\rho\mu}-x^2T_{\mu\nu}, ~ l_\mu{\equiv} ix^\nu
\gamma_\nu s_\mu.
\end{eqnarray}
These three new conserved currents lead to the generators for
dilatation, conformal boost and conformal supersymmetry
transformation.
 However,  the superconformal symmetry may become
 anomalous at quantum level. If all of them, the trace of
 energy-momentum tensor, $T^\mu_{~\mu}$, the $\gamma$-trace
 of supersymmetry current, $\gamma^\mu s_\mu$ and the divergence
 of the chiral $R$-current, $\partial_\mu j^{\mu}$,  get
 contribution from quantum effects,
 they will form a chiral supermultiplet with the $\partial_\mu
j^{\mu}$ playing the role of the lowest component of the
corresponding composite chiral superfield
\cite{anm,sibold,grisaru}.

When considering the ${\cal N}=1$ supersymmetric gauge theory in a
${\cal N}=1$ conformal supergravity background,
  the energy-momentum tensor $T_{\mu\nu}$, the supersymmetry current
 $s_\mu$, and the chiral  (or axial vector)
 $R$-symmetry current $j_\mu$ will
 couple to the gravitational field $g_{\mu\nu}$,
 chiral (or axial) vector field $A_\mu$
 and vector-spinor gravitino field $\psi_{\mu}$ in the
 multiplet of conformal supergravity,
 respectively \cite{fra},
 \begin{eqnarray}
{\cal L}_{\rm ext}=\int d^4x \sqrt{-g}\left( g_{\mu\nu}T^{\mu\nu}
+ A_\mu j^{\mu}+\overline{\psi}_\mu s^{\mu}\right). \label{clag}
\end{eqnarray}
The action (\ref{clag}) shows  that the covariant conservations
of the currents, $\nabla_\mu \theta^{\mu\nu}=D_\mu s^\mu=0$, are
equivalent to the local gauge transformation invariance of the
external supergravity system,
\begin{eqnarray}
\delta g_{\mu\nu} (x) = \nabla_\mu \xi_\nu +\nabla_\nu \xi_\mu
,~~~ \delta\psi_\mu (x)= D_\mu \chi (x).
 \label{egt1}
\end{eqnarray}
Furthermore, the covariant conservation of the chiral (or axial)
vector current $j_\mu $ and
 the vanishing of both the $\gamma$-trace of supersymmetry current
and the trace of energy-momentum  tensor at classical level,
\begin{eqnarray}
\nabla_\mu j^{\mu}=\gamma^\mu s_\mu =T^{\mu}_{~\mu}=0,
\label{supercon}
\end{eqnarray}
mean the Weyl transformation invariance of $g_{\mu\nu}$, the
super-Weyl symmetry and the chiral gauge symmetry of the external
conformal supergravity system,
\begin{eqnarray}
\delta g_{\mu\nu}=g_{\mu\nu} \sigma (x),~~~ \delta \psi_\mu =
\gamma_\mu \eta (x),~~~ \delta A_\mu (x) = \partial_\mu \Lambda
(x). \label{busy}
\end{eqnarray}
This means that the classical superconformal symmetry of the
supersymmetric gauge theory is equivalent to that of the external
conformal supergravity.
 Therefore, in the context of the $AdS/CFT$ (or
more generally gravity/gauge) correspondence the
 superconformal anomaly   in ${\cal N}=1$
supersymmetric gauge theory due to the supergravity external
sources will be reflected in the explicit violations of the bulk
symmetries of ${\cal N}=2$ gauged $AdS_5$ supergravity on the
boundary \cite{witt1,hesk,bian1,chch}.

With no consideration on the quantum correction from the dynamics
of the supersymmetric gauge theory, the external superconformal
anomaly is exhausted  at one-loop level. As given in
Ref.\,\cite{grisaru}, for a general ${\cal N}=1$ supersymmetric
gauge theory with $N_v$ vector and $N_\chi$ chiral multiplets in
an external supergravity background, the chiral R-symmetry and the
Weyl anomalies read
\begin{eqnarray}
\nabla_\mu j^{\mu} &=&\frac{c-a}{24\pi^2} R_{\mu\nu\lambda\rho}
\widetilde{R}^{\mu\nu\lambda\rho}+\frac{5a-3c}{9\pi^2}F_{\mu\nu}
\widetilde{F}_{\mu\nu}, \nonumber\\
T^\mu_{~\mu} &=& \frac{c}{16\pi^2}C_{\mu\nu\lambda\rho}
C^{\mu\nu\lambda\rho}
-\frac{a}{16\pi^2}\widetilde{R}_{\mu\nu\lambda\rho}
\widetilde{R}^{\mu\nu\lambda\rho}+\frac{c}{6\pi^2}
F_{\mu\nu}F^{\mu\nu},\nonumber\\
\gamma_\mu s^\mu &=&\left(A
 R_{\mu\nu\lambda\rho}\gamma^{\lambda\rho}
 +B F_{\mu\nu}\right)D^\mu \psi^\nu.
 \label{sca}
\end{eqnarray}
The coefficients $a$ and $c$ are purely determined by the field
contents.
 For a ${\cal N}=1$ supersymmetric theory in the weak coupling limit,
 they are, respectively, \cite{grisaru}
 \begin{eqnarray}
 c=\frac{1}{24}\left(3 N_v+N_{\chi}\right), ~~~
 a=\frac{1}{48}\left(9 N_v+N_{\chi}\right).
 \label{cc}
 \end{eqnarray}
The coefficients $A$ and $B$ in the super-Weyl anomaly $\gamma_\mu
s^\mu$ are relevant to $a$ and $c$. In above equations,
$\gamma_{\mu\nu}=i/2\,[\gamma_\mu,\gamma_\nu]$;
$F_{\mu\nu}=\partial_\mu A_\nu- \partial_\nu A_\mu$  is the field
strength corresponding to the external $U_R(1)$ vector field
$A_\mu$; $R_{\mu\nu\lambda\rho}$ and $C_{\mu\nu\lambda\rho}$  are
the Riemannian and Weyl tensors corresponding to the gravitational
background field $g_{\mu\nu}$; $\widetilde{R}_{\mu\nu\lambda\rho}$
and $\widetilde{F}_{\mu\nu}$ are the Hodge duals of the Riemannian
tensor and gauge field strength; $D_\mu$ is the covariant
derivative with respect to both the external gravitational and
gauge fields.

\section{ Five-dimensional gauged supergravity and $AdS_5$ boundary
Reduction }

In this section we explain why the superconformal anomaly of a
four-dimensional supersymmetric gauge theory in a conformal
supergravity background can be extracted from a gauged
supergravity in five dimensions \cite{gst,awada,guna2}.

First, all the ${\cal N}=2,4,8$ five-dimensional gauged
supergravities admit an $AdS_5$ classical solution
\cite{gst,awada,guna2},
\begin{eqnarray}
   ds^2 &=& \frac{l^2}{r^2}\left(\eta_{\mu\nu} dx^\mu dx^\nu
   - dr^2\right)
   \label{metrican}
   \end{eqnarray}
all the other fields vanishing. The cosmological term leading
above solution comes from the value of the scalar potential at the
critical point. Further, checking the supersymmetric
transformation of the fermionic field in this background, one can
find the non-vanishing Killing spinor \cite{gst,awada,guna2}. Thus
the $AdS_5$ solution preserve the full supersymmetry of the gauged
supergravity.

Second, we choose this $AdS_5$ classical solution as the vacuum
configuration of the five-dimensional gauged supergravity and
investigate the corresponding dynamical features around such a
vacuum background. For the ${\cal N}=2$ $U(1)$ gauged
supergravity, the  Lagrangian density near the $AdS_5$ vacuum up
to the quadratic terms in spinor fields is of the form \cite{bala}
 \begin{eqnarray}
8\pi G^{(5)}E^{-1} {\cal L} &=& -\frac{1}{2}{\cal R}
-\frac{1}{2}\overline{\Psi}_M^i \Gamma^{MNP}D_N \Psi_{P
i}-\frac{3l^2}{32} {\cal F}_{MN}{\cal F}^{MN}
-\frac{6}{l^2}\nonumber\\
&&-\frac{il^3}{64}E^{-1}\epsilon^{MNPQR} {\cal F}_{MN}{\cal
F}_{PQ}{\cal A}_R
-\frac{3i}{4l}\overline{\Psi}_M^i\Gamma^{MN}\Psi^{N j}
\delta_{ij}\nonumber\\
&&
 -\frac{3il}{32}\left( \overline{\Psi}_M^i\Gamma^{MNPQ}
\Psi_{N i}{\cal F}_{PQ}+2\overline{\Psi}^{M i}\Psi^N_i {\cal
F}_{MN}\right), \label{gaugedfm}
\end{eqnarray}
where ${\cal R}$ is the five-dimensional Riemannian scalar
curvature; $\Psi_M^i$ are the gravitini, $i=1,2$ are the $SU(2)$
$R$ symmetry group indices; ${\cal A}_M$ and ${\cal F}_{MN}$ are
the $U(1)$ gauge field and field strength; $D_M$ is the covariant
derivative with respect to the (modified) spin connection, the
Christoffel and ${\cal A}_M$; $G^{(5)}$ is the five-dimensional
gravitational constant. The supersymmetry transformations at the
leading order in spinor fields read \cite{bala}
\begin{eqnarray}
\delta E_M^{~A}&=&\frac{1}{2}\overline{\cal E}^i\Gamma^A\Psi_{M
i}, ~~~\delta {\cal A}_M = \frac{i}{l}\overline{\Psi}_M^i{\cal
E}_i,
\nonumber\\
\delta \Psi_{M }^i &=& D_{M}{\cal E}^i+ \frac{il}{16}
\left(\Gamma_{M}^{~NP}-4\delta_{M}^{~N} \Gamma^P \right) {\cal
F}_{NP}{\cal E}^i+\frac{i}{2l}\Gamma_M \delta^{ij}{\cal E}_j.
\label{twostm}
 \end{eqnarray}

The investigation on the classical dynamics of the gauged
supergravity around the $AdS_5$ vacuum configuration means looking
for the solution to the classical equation of motion which should
asymptotically approach the $AdS_5$ solution  (\ref{metrican}).
 Geometrically,
this  is actually a process of revealing the asymptotic dynamical
behaviour of the bulk fields near the boundary of $AdS_5$
space-time. The procedure of doing this is straightforward
\cite{nish}. As the first step,  one should perform fix the local
symmetries such as the local Lorentz symmetry, gauge symmetry and
supersymmetry in the radial (fifth)  direction in a way consistent
with the $AdS_5$ classical solution. Then one should reveal the
radial coordinate dependence near the boundary of $AdS_5$
space-time of the solutions to the classical equations of motion
of the fields. There are some delicate points for the fermionic
fields like the gravitino. The fermionic fields are the symplectic
Majorana spinors in  five dimensions, one should show how they
reduce to chiral spinors in four dimensions. Specifically, most of
the fields in the five-dimensional gauged supergravity belong to
certain representation of $R$-symmetry, while the gauged
supergravity in the bulk and the conformal supergravity on the
boundary have different $R$-symmetries. Thus one should exhibit
how the $R$-symmetry in the bulk converts into the one on the
boundary. For the ${\cal N}=2$ gauged supergravity, the solutions
to the classical equations of motion display the follow
leading-order dependence on the radial coordinate near the $AdS_5$
boundary \cite{bala},
  \begin{eqnarray}
{\cal A}_\mu (x,r) &=& {A}_\mu (x)+{\cal O}(r),~~ E_\mu^{~a}
(x,r)=\frac{l}{r} {e}_\mu^{~a}
    (x)+{\cal O}(r ),
   ~~E_r^{~\overline{r}}=\frac{l}{r},\nonumber\\
   \Psi_\mu^R &=& \left(\frac{2l}{r}\right)^{1/2}{\psi}_\mu^R(x),~~
  \Psi_\mu ^L =  \left(2lr\right)^{1/2}{\chi}_\mu^L(x), \nonumber\\
\chi_\mu^L &=& \frac{1}{3}\gamma^\nu\left({D}_\mu\psi_\nu^R
-{D}_\nu\psi_\mu^R\right)-\frac{i}{12}
\epsilon_{\mu\nu}^{~~\lambda\rho} \gamma_5\gamma^\nu
\left({D}_\lambda\psi_\rho^R -{D}_\rho\psi_\lambda^R\right),
\label{refield}
\end{eqnarray}
and all other fields vanish. In above equations,
$\mu,a=0,\cdots,3$ are the Riemannian and local Lorentz indices on
the boundary, respectively, and $\overline{r}$ is the Lorentz
index in the radial direction. The various quantities including
the $\gamma$-matrices and the covariant derivative reduced from
the five-dimensional case are the following \cite{bala},
\begin{eqnarray}
{\gamma}_{a} &=& \Gamma_{a},~~ \Gamma_\mu=
 E_\mu^{~a}\Gamma_{a}=\frac{l}{r}
 {\gamma}_{\mu},
 ~~\Gamma^\mu = \overline{E}^\mu_{~a}\Gamma^{a}
 =\frac{r}{l} {\gamma}^{\mu}, \nonumber\\
 \gamma_5 &=& i\Gamma^{\overline{r}}
 =-i\Gamma_{\overline{r}},~~\gamma_5^2=1; ~~
 {D}_\mu (x) {\equiv}  \nabla_\mu
 +\frac{1}{4}{\omega}_\mu^{~ab}
 {\gamma}_{~ab}-\frac{3}{4}{A}_\mu\gamma_5, \nonumber\\
 \Psi_\mu &\equiv & \Psi_{\mu 1}+i\Psi_{\mu 2}, ~~~
 \Psi^R_\mu\equiv \frac{1}{2} (1-\gamma_5)\Psi_\mu, ~~
 \Psi^L_\mu \equiv \frac{1}{2} (1+\gamma_5).
 \label{rega}
 \end{eqnarray}
Redefining the bulk supersymmetry transformation parameter, ${\cal
E} (x,r)={\cal E}_1(x,r)+i{\cal E}_2(x,r)$, decomposing it into
the chiral components, and further choosing radial coordinate
dependence of ${\cal E}^{L,R}$
 in the same way as
the bulk gravitino,
\begin{eqnarray}
{\cal E}^R(x,r)=\left(\frac{2l}{r}\right)^{1/2}{\epsilon}^R(x),~~~
{\cal E}^L(x,r)= \left(2lr\right)^{1/2} {\eta}^L(x), \label{strp}
\end{eqnarray}
one can find that the bulk supersymmetry transformation reduces to
that  for ${\cal N}=1$ conformal supergravity in four dimensions
with $\epsilon$ and $\eta$ playing the roles of parameters for
supersymmetry and special supersymmetry transformations,
respectively \cite{fra,bala},
\begin{eqnarray}
\delta {e}_\mu^{~a} &=&-\frac{1}{2}
\overline{\psi}_\mu\gamma^{a}{\epsilon},~~
 \delta {\psi}_\mu
= {\nabla}_\mu {\epsilon} -\frac{3}{4}A_\mu \gamma_5
{\epsilon}-\gamma_\mu {\eta},~~ \delta {A}_\mu  =
i\left(\overline{\psi}_\mu\gamma_5{\eta} -\overline{\chi}_\mu
\gamma_5\epsilon \right), \label{desut}
\end{eqnarray}
where all the spinorial quantities, $\psi_\mu (x)$, ${\chi}_\mu
(x)$ ${\epsilon} (x)$ and ${\eta}(x)$
 are Majorana spinors constructed from
their chiral components $\psi^R_\mu (x)$, $\chi^L_\mu (x)$,
${\epsilon}^R (x)$ and ${\eta}^L(x)$.

As for other local symmetries of five-dimensional gauged
supergravity, it has been proved that for any domain wall solution
of the following form which asymptotically approaches the $AdS_5$
solution (\ref{metrican}),
\begin{eqnarray}
ds^2&=& G_{MN} dX^M d X^N=\frac{l^2}{r^2} \left[g_{\mu\nu}(x,r)
dx^\mu dx^\nu - dr^2\right],
   \label{dw1}
 \end{eqnarray}
the diffeormorphism symmetry preserving its above form must be a
combination of the  four-dimensional diffeomorphism symmetry and
the Weyl symmetry \cite{imbi},
\begin{eqnarray}
\delta g_{\mu\nu} (x,r) = 2\sigma (x) \left(
1-\frac{1}{2}r\partial_r\right) g_{\mu\nu}(x,r)+ \nabla_\mu
\xi_\nu (x,r) +\nabla_\nu \xi_\mu (x,r). \label{dedw}
\end{eqnarray}
The $U(1)$ bulk gauge symmetry under the transformation $\delta
{\cal A}_M (x,r)=\partial_M \Lambda (x,r)$, automatically reduces
to the  $U(1)$ chiral (or equivalently axial)
 vector gauge symmetry on the $AdS_5$ boundary.

The above fact indicates that the on-shell five-dimensional gauged
supergravity near the $AdS_5$ vacuum configuration leads  to the
off-shell conformal supergravity on the $AdS_5$ boundary.
Therefore, this has provided the justification that the
superconfromal anomaly of a supersymmetric gauge theory in a
conformal supergravity background can be extracted from the
five-dimensional gauged supergravity.

\section{Holographic Superconformal Anomaly}

\subsection{Holographic Chiral Anomaly}

The holographic origin of the R-symmetry anomaly is the
Chern-Simons (CS) five-form term in the gauge supergravity
\cite{witt1,chu}. For the ${\cal N}=8$ $SO (6)\cong SU(4)$ gauged
supergravity in five dimensions, the CS term is \cite{gst}
\begin{eqnarray}
 S_{\rm CS}[{\cal A}] &=& \frac{l^3}{48 \pi G^{(5)}}\int \mbox{Tr}
 \left[ {\cal A}
(d {\cal A})^2+\frac{3}{2}{\cal A}^3d {\cal A} +\frac{3}{5} {\cal
A}^5\right]
\nonumber\\
&=&\frac{l^3}{48 \pi G^{(5)}}\int \mbox{Tr}\left( {\cal A} {\cal
F}^2-\frac{1}{2}{\cal A}^3 {\cal F}
+\frac{1}{10} {\cal A}^5\right)\nonumber\\
&=& \frac{l^3}{192 \pi G^{(5)}}\int d^5x \epsilon^{MNPQR}
d^{abc}\left({\cal A}_M^a {\cal F}_{NP}^b {\cal F}_{QR}^c- f^{ade}
{\cal A}_M^d {\cal A}_N^e {\cal A}_P^b {\cal F}_{QR}^c
\right.\nonumber\\
&& \left. +\frac{2}{5} f^{ade}f^{efg} {\cal A}_M^d {\cal A}_N^f
{\cal A}_P^g {\cal A}_Q^b {\cal A}_R^c \right),
\end{eqnarray}
where ${\cal A}_M$ and ${\cal F}_{MN}$ are the $SU(4)$ gauge field
and the field strength. A CS term has a particular feature: its
gauge variation is a total derivative. Therefore, under the bulk
gauge transformation,
\begin{eqnarray}
\delta {\cal A}_M^a (x,r)=\left[D_M V(x,r)\right]^a,
\end{eqnarray}
$V(x,r)=V^a(x,r) t^a$ being a gauge transformation parameter, the
other gauge field relevant terms are gauge invariant, but the
gauge transformation of CS term leaves a total derivative term,
 \begin{eqnarray}
\delta_V S_{\rm SUGRA}[{\cal A},\cdots]&=&
\delta_V S_{\rm CS}[{\cal A}]=\int d Q_4^1 (V,{\cal A})\nonumber\\
&=&\frac{l^3}{48\pi G^{(5)}}\int
d\mbox{Tr}\left[{V}\,d\left(AdA+\frac{1}{2} A^3\right)\right].
\label{csgv}
\end{eqnarray}
Choosing the boundary behaviour of the bulk gauge transformation
parameter as the bulk gauge field, $\left.V(x,r)\right|_{r\to 0}=
v(x)$ and making use of the AdS/CFT correspondence (\ref{acc4}),
\begin{eqnarray}
&& \left.\delta_V S_{\rm SUGRA}[{\cal A},\cdots]\right|_{{\cal
A}_M\to A_\mu,\,V\to v } = \delta_v \Gamma_{\rm SYM}
[A_\mu^a,\cdots]= \int d^4x \frac{\delta \Gamma} {\delta
A_\mu^a (x) }\delta A_\mu^a (x)\nonumber\\
&=& \int d^4x j^{a\mu}\delta A_\mu^a (x)
 =\int d^4x j^{a\mu}(x)[D_\mu v(x)]^a
=-\int d^4x v^a (x)[D_\mu j^{\mu} (x)]^a. \label{gtr}
\end{eqnarray}
one can obtain from Eqs.\,(\ref{csgv}) and (\ref{gtr})
\begin{eqnarray}
[D{}^\star j (x)]^a =-\frac{l^3}{48\pi G^{(5)}}\mbox{Tr}t^a\left[
F^2-\frac{1}{2}\left(A^2F+FA^2+AFA\right)+\frac{1}{2} A^4\right].
\end{eqnarray}
Considering the following relations among the $AdS_5$ radius $l$,
string coupling $g_s$, the number $N$ of $D3$-branes, the five-
and ten-dimensional gravitational constants related by the
compactification of the type IIB supergravity on $S^5$ of radius
$l$ \cite{agmo},
\begin{eqnarray}
G^{(5)}=\frac{G^{(10)}}{\mbox{Volume}\,
(S^5)}=\frac{G^{(10)}}{l^5\pi^3}, ~~~ G^{(10)}=8\pi^6g^2_s,
~~~l=\left(4\pi N g_s\right)^{1/4},
\end{eqnarray}
one immediately recognizes the holographic Bardeen (consistent)
anomaly,
\begin{eqnarray}
[D_\mu j^{\mu} (x)]^a =-\frac{N^2}{24\pi^2}e^{-1}
\epsilon^{\mu\nu\lambda\rho}
\partial_\mu \mbox{Tr}t^a\left(A_\nu \partial_\lambda A_\rho+
\frac{1}{2}A_\nu A_\lambda A_\rho \right). \label{nfour}
\end{eqnarray}

For the ${\cal N}=2$ supersymmetric Yang-Mills theory, its
$R$-symmetry group is $U(2)_R{\cong} SU(2)_R\times U(1)_R$. It is
the $U(1)_R$ that becomes anomalous. The dual gravitational theory
is the five-dimensional $SU(2)\times U(1)$ gauged ${\cal N}=4$
supergravity. The holographic chiral $U(1)_R$ anomaly comes from
the $SU(2)\times U(1)$-mixed CS term in the gauged supergravity
\cite{awada,ferr},
\begin{eqnarray}
S_{\rm CS}[{\cal W}, {\cal A},\cdots] &=& \frac{l^3}{64\pi
G^{(5)}} \int \mbox{Tr}\left({\cal G}\wedge {\cal G}\right)\wedge
{\cal A},
\end{eqnarray}
where ${\cal W}$ and  ${\cal A}$ are the $SU(2)$ and $U(1)$ gauge
fields, and ${\cal G}$ the $SU(2)$ field strength. The reduction
of the bulk $U(1)$ gauge transformation $\delta {\cal A}=d V$ to
the $AdS_5$ boundary leads to the holographic $U_R(1)$ anomaly,
\begin{eqnarray}
\partial_\mu \left(e j^\mu\right)=-\frac{N^2}{32\pi^2}
\epsilon^{\mu\nu\lambda\rho} \mbox{Tr}\left( G_{\mu\nu}
G_{\lambda\rho}\right). \label{ntwo}
\end{eqnarray}
The justification that the boundary value $A_\mu(x)$ of the bulk
gauge field ${\cal A}_{M}$ is considered as the external chiral
(or axial) gauge field in four dimensions is implied from the
boundary reductions of the bulk covariant derivative and of the
supersymmetric transformation listed in Eqs.\,(\ref{rega}) and
(\ref{desut}).

However, Eqs.\,(\ref{nfour}) and (\ref{ntwo}) do not contain the
gravitational background contribution shown in the general
expression (\ref{sca}). In fact, for ${\cal N}=4$ supersymmetric
Yang-Mills theory, there exists no gravitational contribution.
 The reason is that
 the field contents of ${\cal N}=4$ SYM
 can be considered as one ${\cal N}=1$ vector multiplet
 plus three chiral multiplets in the adjoint representation
 of $SU(N)$ and according to Eq.\,(\ref{cc}) this yields $c=a=(N^2-1)/4$ \cite{blau}.
Thus the CS term composed of the $SU_R(4)$ gauge field is fully
responsible for the holographic source of the chiral R-symmetry
anomaly.
 For the general ${\cal N}=1,2$ supersymmetric gauge theories, Eq.\,(\ref{cc})
 shows that usually $a\neq c$, hence the
 gravitational background contribution to the chiral anomaly
  should arise. Its absence implies that the five-dimensional
  gauged supergravity (or the type IIB supergravity
  in $AdS_5\times X^5$ background)
  is only the lowest order approximation to type IIB superstring theory)
  and corresponds only to the leading order of the large-$N$ expansion
 of supersymmetric gauge theory.
   In Ref.\,\cite{ahar}
  it was shown that for an ${\cal N}=2$ supersymmetric $USp(2N)$
  gauge theory coupled to two hypermultiplets in the fundamental and
  antisymmetric tensor representations of the gauge group, respectively,
  the gravitational background part of the holographic chiral anomaly
  does originate from a mixed CS term. However, this CS
  terms is obtained from the compactification on $S^3$ of the Wess-Zumino
  term describing the interaction of the R-R 4-form field with
  eight $D7$-branes and one orientifold 7-plane system. Specifically,
  this gravitational background term is at the subleading $N$-order
  rather than the leading $N^2$-order in the large-$N$ expansion
  of the ${\cal N}=2$ supersymmetric $USp(2N)$ gauge theory. This fact
   exposes the limitation of the gauged supergravity in
   providing an equivalent physical description
  to the supersymmetric gauge theory.

\subsection{Holographic Weyl Anomaly}

 The holographic
origin of the Weyl anomaly of a supersymmetric gauge theory
 lies in the $AdS_5$ boundary behaviour of the on-shell action
of the gauged  supergravity. Due to the infinity of the boundary,
the on-shell action of the five-dimensional gauged supergravity in
$AdS_5$ background suffers from the infrared divergences when
approaching the boundary. Therefore, one must perform a so-called
`` holomorphic renormalization'' \cite{bian1}. That is, first
 introducing an IR cut-off when one integrate over the radial (fifth)
 coordinate to evaluate the on-shell action, then similar
 to dealing with the UV divergence in a renormalizable
 quantum field theory,  defining a counterterm according to
a preferred renormalization condition to cancel the IR divergence,
finally removing the cut-off to get
 the renormalized on-shell action for the gauged supergravity.
Specifically, Eq.\,(\ref{dedw}) shows that the bulk diffeomorphism
symmetry of the gauged supergravity decomposes into the
diffeomorphism symmetry
 and the Weyl symmetry on the
boundary \cite{imbi}. These two symmetries cannot be preserved
simultaneously in implementing
 the holomorphic renormalization. Thus
if one requires the diffeomorphism symmetry preserved,
 the holographic Weyl
anomaly of a supersymmetric gauge theory will arise.

We take the ${\cal N}=8$ $SO(6)$ gauged supergravity in five
dimensions as an example, and choose the truncated action
consisting only of the Einstein-Hilbert action and the
non-vanishing scalar potential \cite{hesk,bian1},
\begin{eqnarray}
8\pi G^{(5)}E^{-1} {\cal L}_{\rm trunc} =-\frac{1}{2}{\cal
R}-P[\phi]. \label{truaction}
\end{eqnarray}
The corresponding Einstein equation is
\begin{eqnarray}
{\cal R}_{MN}-\frac{1}{2} {\cal R} G_{MN}=P[\phi=0] G_{MN}.
\label{ee}
\end {eqnarray}
The solution to the Einstein equation (\ref{ee}) is the domain
wall (\ref{dw1}). It should be emphasized that the existence of
scalar field is necessary for the domain wall solution
(\ref{dw1}). Otherwise there will be no non-trivial vacuum
configurations and the domain wall solution  does not exist. Near
the $AdS_5$ boundary ($r\rightarrow 0$), the solution $g_{\mu\nu}
(x,r)$ admits the following expansion \cite{hesk,bian1},
\begin{eqnarray}
g_{\mu\nu}(x,r)&=&
g_{(0)\mu\nu}(x)+g_{(2)\mu\nu}(x)\frac{r^2}{l^2}
\nonumber\\
&&+ \left[g_{(4)\mu\nu}+ h_{1(4)\mu\nu}\ln \frac{r^2}{l^2}
+h_{2(4)\mu\nu}\left(\ln \frac{r^2}{l^2}\right)^2\right]
\left(\frac{r^2}{l^2}\right)^2 +\cdots. \label{expa}
\end{eqnarray}
Substituting (\ref{expa}) into the Einstein equation (\ref{ee})
with the cosmological constant provided by the value of the scalar
potential at the critical point $\phi=0$, one can determine the
coefficients $g_{(2k)\mu\nu}$, $h_{\mu\nu}$ in terms of
$g_{(0)\mu\nu}$ \cite{hesk,bian1},
\begin{eqnarray}
g_{(2)\mu\nu} &=&\frac{l^2}{2}\left( R_{\mu\nu}-\frac{1}{6}R
g_{(0)\mu\nu}
\right),\nonumber\\
h_{1(4)\mu\nu}
&=&\frac{l^4}{8}\left(R_{\mu\lambda\nu\rho}R^{\lambda\rho}
+\frac{1}{6}\nabla_\mu\nabla_\nu R-
\frac{1}{2}\nabla^2R_{\mu\nu}-\frac{1}{3}R R_{\mu\nu}\right)\nonumber\\
&&+\frac{l^4}{32}g_{(0)\mu\nu}\left(\frac{1}{3}\nabla^2
R+\frac{1}{3} R^2
-R_{\lambda\rho}R^{\lambda\rho}\right), \nonumber\\
h_{2(4)\mu\nu} &=& 0,~~~\mbox{Tr}g_{(4)}
=\frac{1}{4}\mbox{Tr}\left(g_{(2)}\right)^2,
\nonumber\\
\nabla^\nu g_{(4)\mu\nu} &=& \nabla^\nu\left\{
-\frac{1}{8}\left[\mbox{Tr}g_{(2)}^2-\left(\mbox{Tr}g_{(2)}\right)^2
\right] g_{(0)\mu\nu}+\frac{1}{2}\left(g_{(2)}^2\right)_{\mu\nu}
-\frac{1}{4}g_{(2)\mu\nu}\mbox{Tr} g_{(2)}\right\},
\end{eqnarray}
where $\mbox{Tr}g_{(2)}=g_{(0)}^{\mu\nu}g_{(2)\mu\nu}$,
$R_{\mu\nu\lambda\rho}$, $R_{\mu\nu}$ and $R$ are the Riemannian
tensor, Ricci tensor and curvature scalar corresponding to
$g_{(0)\mu\nu}$, respectively. Inserting above solution into
Eq.\,(\ref{truaction}) to evaluate the on-shell action, one finds
that it is divergent when approaching the $AdS_5$ boundary. Thus
one must perform the radial integration in a finite domain by
introducing a cut-off $r=\epsilon >0$ and get the regularized
on-shell action,
\begin{eqnarray}
S_{\rm reg} &=& \frac{1}{8\pi G^{(5)}}\int_{r=\epsilon >0} d^5X
E(x,r)\left(-\frac{1}{2}{\cal R}-P[\phi=0]\right)\nonumber\\
&=& \frac{l^5}{8\pi G^{(5)}}\int_{\epsilon} \frac{dr}{r^5}\int
 d^4x \sqrt{g (x,r)}\,
\frac{2}{3}P[\phi=0]\nonumber\\
&=& -\frac{l^5}{8\pi G^{(5)}}\int_{\epsilon} \frac{dr}{r^5}\int
 d^4x \sqrt{g (x,r)}\,
\frac{4}{3}\Lambda = \frac{l^3}{\pi G^{(5)}}\int_{\epsilon}
\frac{dr}{r^5}\int
 d^4x \sqrt{g (x,r)}\nonumber\\
 &=&\frac{l^3}{\pi G^{(5)}}\int
 d^4x \sqrt{g_0(x)} \int_{\epsilon} \frac{dr}{r^5}
 \left\{1+\frac{r^2}{2l^2}\mbox{Tr}g_{(2)}
 +\frac{r^4}{4l^4}\left[\left( \mbox{Tr}g_{(2)}\right)^2
 -\mbox{Tr}g^2_{(2)}\right]+\cdots\right\}\nonumber\\
&=&\frac{l^3}{\pi G^{(5)}}
 \int d^4x\sqrt{g_{(0)}}\left[-\frac{1}{4\epsilon^4}
 -\frac{1}{12\epsilon^2}R
-\frac{1}{16}\ln\frac{\epsilon}{l}\,\left(R_{\mu\nu}R^{\mu\nu}
-\frac{1}{3}R^2 \right) +{\cal L}_{\rm
finite}\right]\nonumber\\
&=& S_{\epsilon^{-4}}+S_{\epsilon^{-2}}+S_{\ln\epsilon}+S_{\rm
finite}. \label{regaction}
\end{eqnarray}
In writing down (\ref{regaction}) the Einstein equation
(\ref{ee}), the identification $\Lambda=-2P[\phi=0]=-{6}/{l^2}$
and the matrix operation
\begin{eqnarray}
\sqrt{\det (1+A)}&=&\exp\left[\frac{1}{2}\mbox{Tr} \ln
\left(1+A\right)\right] =\exp\left[\frac{1}{2}\mbox{Tr} \left(A-
\frac{1}{2}A^2+\cdots \right)\right]
\nonumber\\
&=& 1+\frac{1}{2}\mbox{Tr}A+\frac{1}{4} \left[
\left(\mbox{Tr}A\right)^2 -\mbox{Tr}A^2\right]+\cdots,
\end{eqnarray}
have been employed. ${\cal L}_{\rm fin}$ consists of the terms
standing the $\epsilon{\rightarrow}0$ limit.

To get the renormalized on-shell action action, one must first
define a subtracted action  by introducing the counterterms to
cancel the IR divergence in the limit $\epsilon\rightarrow 0$,
\begin{eqnarray}
S_{\rm sub}[g_{(0)\mu\nu},\epsilon^2/l^2,\cdots]=S_{\rm reg}.
+S_{\rm counter}
\end{eqnarray}
A holographically renormalized on-shell action is yielded after
removing the regulator,
\begin{eqnarray}
S_{\rm
ren}[g_{(0)\mu\nu},\cdots]=\lim_{\epsilon{\rightarrow}0}S_{\rm
sub}[g_{(0)\mu\nu},\epsilon^2/l^2,\cdots]
\end{eqnarray}
In adding the counterterm, there arise a finite ambiguity similar
to cancelling the UV divergence in a perturbative quantum field
theory. This ambiguity can be fixed by the symmetry requirement.
According to Eq.\,(\ref{dedw}), the bulk diffeomorphism
transformation converts into  a four-dimensional Weyl and a
diffeormorphism transformations near the $AdS_5$ boundary.
Requiring the four-dimensional diffeomorphism symmetry
 preserved in performing subtraction, one can introduce
 the following counterterm similar to the minimal subtraction
 in the dimensional regularization of
  the perturbative quantum field theory \cite{hesk,bian1},
 \begin{eqnarray}
 S_{\rm counter} &=& \frac{l^3}{\pi G^{(5)}}
 \int d^4x\sqrt{g_{(0)}}\left[\frac{1}{4\epsilon^4}
 +\frac{1}{12\epsilon^2}R
+\frac{1}{16} \left(R_{\mu\nu}R^{\mu\nu} -\frac{1}{3}R^2
\right)\ln\frac{\epsilon}{l}\right],
 \end{eqnarray}
 and consequently, the renormalized on-shell action is just
 the finite part of the regularized one,
\begin{eqnarray}
S_{\rm ren} &=& \frac{l^3}{\pi G^{(5)}} \int d^4x
\sqrt{g_{(0)}}{\cal L}_{\rm finite}.
\end{eqnarray}
There are several ways to extract the Weyl anomaly from the
renormalized action \cite{hesk,bian1}.
 The most straightforward way is
considering the scale transformation of the regularized action,
i.e., choosing the parameter $\sigma (x)$ of the Weyl
transformation as a constant $\sigma$. The regularized action is
invariant under the combination of two transformations, $\delta
g_{(0)\mu\nu}=2\sigma g_{(0)\mu\nu}$, $\delta \epsilon=2 \sigma
\epsilon$. That is \cite{hesk},
\begin{eqnarray}
\left(\delta_{g_0}+\delta_\epsilon\right) S_{\rm reg}
=\left(\delta_{g_0}+\delta_\epsilon\right) \left(
S_{\epsilon^{-4}}+S_{\epsilon^{-2}}+S_{\ln\epsilon}+S_{\rm
finite}\right) =0.
\end{eqnarray}
However, it has been found that \cite{hesk}
\begin{eqnarray}
\delta_{g_0} \left(S_{\epsilon^{-4}}+S_{\epsilon^{-2}}\right)
=0,~~~ \delta_\epsilon \left(
S_{\epsilon^{-4}}+S_{\epsilon^{-2}}\right) =0,~~~ \delta_{g_0}
S_{\ln\epsilon}=0, ~~~ \delta_\epsilon S_{\rm finite} =0.
\end{eqnarray}
This leads to
\begin{eqnarray}
\delta_{g_0} S_{\rm finite}&=&\delta_{g_0} S_{\rm ren}=\int d^4x
\sqrt{g_{(0)}}\langle T^{\mu}_{~\mu}\rangle\sigma
=-\delta_\epsilon S_{\ln\epsilon}\nonumber\\
&=&\frac{l^3}{8\pi G^{(5)}}\int d^4x \sqrt{g_{(0)}}
\left(R_{\mu\nu}R^{\mu\nu}-\frac{1}{3}R^2 \right)\sigma,
\end{eqnarray}
and yields the Weyl anomaly in the gravitational background
\cite{hesk,bian1},
\begin{eqnarray}
\langle T^\mu_{~\mu}\rangle =
\frac{N^2}{4\pi^2}\left(R_{\mu\nu}R^{\mu\nu}-\frac{1}{3}R^2\right).
\label{trace}
\end{eqnarray}
It can be further rewritten as the combination of the $A$- and
$B$-type anomalies, i.e, the sum of the Euler number density and
the square of the Weyl tensor \cite{hesk,deser},
\begin{eqnarray}
\langle T^\mu_{~\mu}\rangle &=& -\frac{N^2}{\pi^2} \left(E_4+W_4\right), \nonumber\\
E_4 &=&
\frac{1}{8}\widetilde{R}_{\mu\nu\lambda\rho}\widetilde{R}^{\mu\nu\lambda\rho}
=\frac{1}{8}\left( R^{\mu\nu\lambda\rho}R_{\mu\nu\lambda\rho}
-4R^{\mu\nu}R_{\nu\nu}+R^2 \right),\nonumber\\
W_{4} &=& -\frac{1}{8}
{C}_{\mu\nu\lambda\rho}{C}^{\mu\nu\lambda\rho} =-\frac{1}{8}\left(
R^{\mu\nu\lambda\rho}R_{\mu\nu\lambda\rho}
-2R^{\mu\nu}R_{\mu\nu}+\frac{1}{3}R^2 \right).
\end{eqnarray}
It is the trace anomaly  of the ${\cal N}=4$ supersymmetric
Yang-Mills theory in the external gravitational field  at the
leading order of large-$N$ expansion \cite{hesk}.

The gauge field contribution to the Weyl anomaly can be similarly
calculated when switching on the gauge field sector of the ${\cal
N}=8$ gauged supergravity in five dimensions \cite{bian1}. If one
considers only the bilinear terms in the gauge fields, the $SO(6)$
gauge field can be approximated by some uncoupled Abelian sectors
${\cal A}_M$ \cite{bian2}. The solution to the gauge field
equation near the $AdS_5$ boundary is \cite{bian1}
\begin{eqnarray}
{\cal A}_\mu (x,r)= A_\mu (x)+\frac{r^2}{l^2}\left[ A_{(2)\mu}(x)+
\widetilde{A}_{(2)\mu}(x) \ln \frac{r^2}{l^2}\right]+\cdots,
\label{ge}
\end{eqnarray}
where $A_{(2)\mu}(x)$ and $\widetilde{A}_{(2)\mu}(x)$ can be
expressed as the functional of $A_\mu (x)$ when inserting
Eq.\,(\ref{ge}) into the classical equation of motion for ${\cal
A}_\mu$. The regularized action of the gauge field sector is
\cite{bian1}
\begin{eqnarray}
S_{\rm reg}=\frac{l^3}{8\pi G^{(5)}}\int d^4x \sqrt{g_{(0)}}
\left[\left(-\frac{1}{4} F_{\mu\nu}F^{\mu\nu}\right)\ln
\frac{\epsilon}{l} +{\cal L}_{\rm finite}\right]
\end{eqnarray}
The gauge field part of the Weyl anomaly given in Eq.\,(\ref{sca})
can be derived in the same way as the gravitational case.

\subsection{Holographic Supersymmetry Current Anomaly}

The production of the super-Weyl anomaly of a four-dimensional
 supersymmetric gauge theory from the five-dimensional
 gauged supergravity
 lies in two aspects. First,
as a supersymmetric theory, the supersymmetry transformation of
the Lagrangian of the gauged supergravity must be composed of the
total derivative terms. These terms cannot be naively ignored due
to the existence of the boundary $AdS_5$. Second, near the $AdS_5$
boundary the bulk supersymmetry transformation  decomposes into
the four-dimensional supersymmetry and super-Weyl transformations
as shown in Eq.\,(\ref{desut}). If we require four-dimensional
supersymmetry preserved on the boundary, the total derivative
terms should yield the anomaly of the supersymmetry current via
the AdS/CFT correspondence. Therefore, the key point is to
calculate the supersymmetric
 variation of the gauged supergravity
 and get  the total derivative terms. Then
 putting these terms on the $AdS_5$ boundary, one should find
  the holographic supersymmetry current anomaly.

We have worked out these total derivative terms of the simplest
case in the five dimensional gauged supergravities, the ${\cal
N}=2$ $U(1)$ gauged supergravity whose Lagrangian is given in
Eq.\,(\ref{gaugedfm}). The concrete calculation is very lengthy
and has been displayed in a great detail in Ref.\,\cite{chch}. The
variation of the Lagrangian (\ref{gaugedfm}) under the
supersymmetric transformation (\ref{twostm}) yields
\begin{eqnarray}
\delta S &=& \frac{1}{8\pi G^{(5)}}\int d^5x E \nabla_M \left(
-\frac{9il}{16}
 \overline{\cal E}^i\Psi_{N i}{\cal F}^{MN}
 -\frac{1}{2}\overline{\cal E}^i\Gamma^{MNP}\nabla_N
 \Psi_{P i}\right.\nonumber\\
 && + \frac{3}{8}
  \overline{\cal E}^i\Gamma^{MNP}
 \Psi_{P}^j\delta_{ij}{\cal A}_N
 - \frac{3il}{32}E^{-1}\epsilon^{MNPQR}
 \overline{\cal E}^i\Gamma_{R}
 \Psi_{N i}{\cal F}_{PQ}\nonumber\\
 &&\left.+\frac{9}{4}\overline{\cal E}^i\Gamma^{MN}
 \Psi_{N}^i\delta_{ij}
 +\frac{l^2}{16}E^{-1}
 \epsilon^{MNPQR}
 \overline{\cal E}^i \Psi_{R}{\cal A}_N {\cal F}_{PQ}\right).
\label{var3}
\end{eqnarray}
In deriving above derivative terms, one must bear in mind that in
the Lagrangian (\ref{gaugedfm}) the $U(1)$ gauge field is
imaginary and
 the gravitino field is an $SU(2)$ symplectic Majorana spinor,
 \begin{eqnarray}
{\cal A}_M^\star &=& -{\cal A}_M, ~~~ \Psi^i =
C^{-1}\Omega^{ij}\overline{\Psi}_j^T=C^{-1}\overline{\Psi}^{iT},
 ~~~\overline{\Psi}^i=-\Psi^{iT}C,\nonumber\\
&&   \overline{\Psi}^i\Gamma_{M_1\cdots M_n}\Phi_i =
 -\Psi^{iT}C\Gamma_{M_1\cdots M_n}C^{-1}\overline{\Phi}_i^T\nonumber\\
&=&\left\{\begin{array}{l}
 -\Psi^{iT}\Gamma_{M_1\cdots M_n}\overline{\Phi}^T_i
=\overline{\Phi}_i\Gamma_{M_1\cdots M_n} \Psi^i
=-\overline{\Phi}^i\Gamma_{M_1\cdots M_n} \Psi_i,~~ n=0,1,4,5,\\
\Psi^{iT}\Gamma_{M_1\cdots M_n}\overline{\Phi}^T_i
=-\overline{\Phi}_i\Gamma_{M_1\cdots M_n} \Psi^i
=\overline{\Phi}^i\Gamma_{M_1\cdots M_n} \Psi_i,~~ n=2,3,
\end{array} \right.\, .
\end{eqnarray}
The convention for  the $\Gamma$-matrix in five dimensions and the
frequently used relations are chosen as the following,
\begin{eqnarray}
\Gamma_{MN}&=&\frac{1}{2}[\Gamma_M,\Gamma_N], ~~
\Gamma^{MNP}=-\frac{1}{2!}E^{-1}\,
\epsilon^{MNPQR}\Gamma_{QR},\nonumber\\
\Gamma^{MNPQ}&=& E^{-1}\, \epsilon^{MNPQR}\Gamma_{R},
~~\Gamma_{MNPQR}=E\, \epsilon_{MNPQR}.
\nonumber \\
 \Gamma_{MN}\Gamma_{PQ}
&=& E\,\epsilon_{MNPQR}\Gamma^R- \left(G_{MP}
G_{NQ}-G_{MQ}G_{NP}\right),
\nonumber\\
\Gamma_{M}\Gamma_{NP} &=& \Gamma_{MNP}
+G_{MN}\Gamma_P-G_{MP}\Gamma_N, \nonumber\\
\Gamma^{MNP}\nabla_N \nabla_P \Psi_i &=&
\frac{1}{2}\Gamma^{MNP}[\nabla_N, \nabla_P] \Psi_i =
\frac{1}{8}\Gamma^{MNP}{\cal R}_{NP AB}\Gamma^{AB}\Psi_i.
 \end{eqnarray}
 The following Ricci and Bianchi identities for
 the Riemannian curvature tensor and the $U(1)$ field strength are
 employed in the calculation,
 \begin{eqnarray}
 \epsilon^{MNPQR} {\cal R}_{SPQR}=0, ~~~
 \epsilon^{MNPQR}\nabla_N {\cal R}_{STPQ}=0,~~~
 \epsilon^{MNPQR}\nabla_N {\cal F}_{QR}=0.
 \end{eqnarray}
 Specifically, due to the nocommutativity between $\nabla_M$ and
 $\Gamma_{M_1\cdots M_n}$, we reiteratively make the operation,
 \begin{eqnarray}
 \Gamma_{M_1\cdots M_n} \nabla _M (\cdots)
 &=&\left[\Gamma_{M_1\cdots M_n},\nabla _M\right] (\cdots)
 +\nabla_M \left[\Gamma_{M_1\cdots M_n} (\cdots)\right].
 \end{eqnarray}
In this case it is convenient to choose the inertial coordinate
system,
 i.e. the Christoffel symbol $\Gamma^{M}_{~NP}=0$. Consequently,
 the metricity condition leads to
 $\partial_M E_N^{~A}=0$,
 and hence the modified spin connection $\Omega_{M AB}$
  contains only the quadratic fermionic terms. This simplifies the
  calculation greatly since we retain only the quadratic fermionic terms
  in calculating the supersymmetry variation.

 The holographic super-Weyl anomaly can be extracted from above
 total derivative terms. First, we take into account the radial
 coordinate dependence of bulk fields
 and of the supersymmetry transformation parameter ${\cal E}^i$
 near the $AdS_5$ boundary
 as well as the connection between five- and four-dimensional
 $\gamma$-matrices listed in
 (\ref{refield}), (\ref{rega})
 and (\ref{strp}). Second, to avoid the possible IR divergence due to
 the infinite $AdS_5$ boundary,  we must integrate over
 the radial coordinate to the cut-off $r=\epsilon$ and then
 take the limit $\epsilon\to 0$. Finally, we use the fact that
 the metric on the boundary should be the induced metric \cite{hesk,bian1}
 \begin{eqnarray}
 {g}_{\mu\nu}(x)=\left.\frac{l^2}{\epsilon^2}{g}_{\mu\nu}(x,\epsilon)
 \right|_{\epsilon\to 0}
\end{eqnarray}
rather than ${g}_{\mu\nu}(x,\epsilon)$ \cite{bian1}. With all
these considerations together, we have found
  that the non-vanishing contribution comes only from the
  term $E^{-1}\epsilon^{MNPQR}\overline{\cal E}^i\Gamma_{R}
 \Psi_{N i}{\cal F}_{PQ}$. Therefore, we obtain \cite{chch}
 \begin{eqnarray}
 \delta S &=& \frac{3il^3}{8\times 32\pi G^{(5)}}
 \int d^4 x \epsilon^{\mu\nu\lambda\rho}
 F_{\nu\lambda}\overline{\eta}\gamma_\rho\chi_\mu,
 \label{fvar}
  \end{eqnarray}
  where ${\chi}_\mu$ is the Majorana spinor constructed from the
  left-handed spinor $\chi_\mu^L$ given in (\ref{refield}).

Eq.\,(\ref{fvar}) definitely leads to the super-Weyl anomaly in
the context of AdS/CFT correspondence (\ref{acc4}) since it is
proportional to the special supersymmetry transformation parameter
$\eta$. Inserting the explicit form $\chi_\mu$ expressed in terms
of the gravitino $\psi_\mu$ in four-dimensional ${\cal N}=1$
conformal supergravity, we have
 \begin{eqnarray}
 \delta S &=& \int d^4x \overline{\eta}\gamma^\mu s_\mu \nonumber\\
 &=& \frac{3il^3}{8\times 32\pi G^{(5)}}\int d^4 x
 \epsilon^{\mu\nu\lambda\rho}
 F_{\nu\lambda}\overline{\eta}\gamma_5\gamma_\rho \gamma^\alpha
 \left[ \frac{1}{3} \left(D_\mu\psi_\alpha- D_\mu\psi_\alpha\right)
 -\frac{i}{6}\epsilon_{\mu\alpha\sigma\delta}\gamma_5 D^\sigma\psi_\delta
  \right]\nonumber\\
  &=& -\frac{l^3}{8\times 16\pi G^{(5)}}\int d^4x
  \left[F^{\mu\nu}D_\mu \psi_\nu
  +\epsilon^{\mu\nu\lambda\rho}\gamma_5 F_{\mu\nu} D_\lambda\psi_\rho
  \right.\nonumber\\
  &&\left.+\frac{1}{2}\sigma^{\mu\nu} F_{\nu\lambda}
  \left(D_\mu\psi^\lambda-D^\lambda\psi_\mu\right)\right],
 \end{eqnarray}
 where we have used the $\gamma$-matrix algebraic relations,
 $\gamma^\mu\gamma^\nu=g^{\mu\nu}-i\gamma^{\mu\nu}$,
 $\gamma_5\gamma^{\mu\nu}={i}\epsilon^{\mu\nu\lambda\rho}
 \gamma_{\lambda\rho}/{2}$.
The gauge field part of the holographic super-Weyl anomaly of the
$SU(N)$ supersymmetric gauge theory at the leading-order of the
large-$N$ expansion is thus yielded,
\begin{eqnarray}
\gamma_\mu s^\mu &=& \frac{N^2}{64\pi^2}\left[F^{\mu\nu}D_\mu
\psi_\nu
  +\epsilon^{\mu\nu\lambda\rho}\gamma_5 F_{\mu\nu} D_\lambda\psi_\rho
  +\frac{1}{2}\sigma^{\mu\nu} F_{\nu\lambda}
  \left(D_\mu\psi^\lambda-D^\lambda\psi_\mu\right)\right].
\label{gtra}
\end{eqnarray}

There should also has a contribution from
 the external gravitational background shown in Eq.\,(\ref{sca}),
 which was found long time ago \cite{abb}.
The reason for having failed to reproduce the
 gravitational part is not clear to us yet,
we have the following two speculations based on the process of
deriving the gravitational background parts in both the
holographic Weyl and chiral anomalies \cite{blau,ahar}.

The first intuitive argument, as mentioned before, is that the
five-dimensional gauged supergravity (or the type IIB supergravity
in $AdS_5\times X^5$ background) is only the lowest approximation
to the type IIB superstring theory in $AdS_5\times X^5$
background. Thus, it is possible that the gravitational part
cannot be revealed within the five-dimensional gauged supergravity
itself, and one must consider the higher-order gravitational
action
 such as the Gauss-Bonnet term generated from the superstring theory
 \cite{berg2}. The supersymmetry variation of the gauged supergravity
 containing the
 high-order gravitational term and the corresponding fermionic
 terms required by supersymmetry may lead to the gravitational contribution
 to the super-Weyl anomaly.

The other possible reason for the failure of getting the
 gravitational background contribution is
that in Eq.\,(\ref{refield}) only the leading-order of radial
coordinate dependence of the bulk fields near the $AdS_5$ boundary
is taken into account. As shown above in deriving the holographic
Weyl anomaly,
 when one makes a complete near-boundary
analysis and considers the asymptotic expansion of the bulk fields
in terms of the radial coordinate beyond the leading-order until
the emergence  of the logarithmic term \cite{hesk,bian1}, the
higher-order gravitational terms can appear in the on-shell action
 \cite{hesk}, and they lead to the holographic Weyl anomaly composed
of the $R_{\mu\nu} R^{\mu\nu}$ and $R^2$ terms. Therefore, it is
also possible that  the gravitational background part in the
super-Weyl anomaly can arise if one takes into account the
logarithmic term in the expansion of the on-shell bulk fields. In
this case the on-shell action of the five-dimensional gauged
supergravity should have the infrared divergence when approaching
 the $AdS_5$ boundary. One must perform the holographic renormalization
 to get the renormalized on-shell action \cite{bian1}.

 We have not realized whether there are any
 physical reasons for the difference between these two holographic
  contributions to the super-Weyl anomaly. The essence of
  the holographic anomaly
  is the anomaly inflow from the bulk theory to the $AdS_5$ boundary \cite{callan}.
  Thus the absence of the gravitational part
   might be relevant to the difference between the
  anomaly inflows contributed from the gravitational and gauge
  background fields.

\section{Summary}

We have reviewed how the superconfromal anomaly multiplet of a
supersymmetric gauge theory in a conformal supergravity background
can be produced via the AdS/CFT correspondence. The type IIB
supergravity in $AdS_5\times X^5$ background
 reduce to a gauged supergravity in five dimensions since
such a background provides a compactification on $X^5$, thus the
AdS/CFT correspondence implies that there should exist a
holographic correspondence between the gauged supergravity in five
dimensions and a four-dimensional $SU(N)$ supersymmetric gauge
theory in certain phase at the large-$N$ limit. Based on this
consideration, we make use of the fact that the five-dimensional
gauged supergravity admits a classical $AdS_5$ solution preserving
the full supersymmetry. Then it is found that around this $AdS_5$
vacuum configuration the supermultiplet of the on-shell
five-dimensional gauged supergravity converts into the
 off-shell conformal supergravity multiplet in four dimensions.
 Therefore, the holographic relation
between  the $AdS_5$ gauged supergravity and the four-dimensional
 supersymmetric $SU(N)$ gauge theory at the large-$N$ limit can be
established in the sense that the conformal supergravity on one
hand has furnished a background for a classical superconformal
gauge theory and on the other hand is a $AdS_5$ boundary theory
for the five-dimensional gauged supergravity. Therefore, it is
natural to reproduce the superconformal anomaly of a
four-dimensional supersymmetric gauge theory from the $AdS_5$
gauged supergravity. But still it is amazing that these three
distinct anomalies can be extracted in the framework of a
five-dimensional gauged supergravity.

\vspace{3ex}

\noindent{\large \bf Acknowledgments:}
 This work is supported by the Academy of Finland under
 the Project No.\,54023.

\end{document}